\documentclass[11pt,preprint,graphicx]{aastex631}
\usepackage{newtxtext,newtxmath}
\usepackage[T1]{fontenc}
\usepackage{ae,aecompl}
\usepackage{graphicx}
\usepackage{threeparttable}

\newcommand{\etal}{\it et al. \rm }

\begin{document}

\title{Stellar Mass-to-light Ratios: Composite Bulge+Disk Models and the Baryonic Tully-Fisher
Relation} 

\author[0000-0003-2022-1911]{James Schombert}
\affiliation{Institute of Fundamental Physics, University of Oregon, Eugene, OR 97403}
\author[0000-0002-9762-0980]{Stacy McGaugh}
\affiliation{Department of Astronomy, Case Western Reserve University, Cleveland, OH 44106}
\author[0000-0002-9024-9883]{Federico Lelli}
\affiliation{Arcetri Astrophysical Observatory (INAF), Florence, Tuscany, IT}

%\date{}
%\pubyear{2022}
%\label{firstpage}
%\pagerange{\pageref{firstpage}--\pageref{lastpage}}
%\maketitle

\begin{abstract}

\noindent We present stellar population models to calculate the mass-to-light ratio
($\Upsilon_*$) based on galaxy's colors ranging from $GALEX$ FUV to {\it Spitzer}
IRAC1 at 3.6$\mu$m.  We present a new composite bulge+disk $\Upsilon_*$ model that
considers the varying contribution from bulges and disks based on their optical and
near-IR colors.  Using these colors, we build plausible star formation histories and
chemical enrichment scenarios based on the star formation rate-stellar mass and
mass-metallicity correlations for star-forming galaxies.  The most accurate
prescription is to use the actual colors for the bulge and disk components to
constrain $\Upsilon_*$; however, a reasonable bulge+disk model plus total color only
introduces 5\% more uncertainty.  Full bulge+disk $\Upsilon_*$ prescriptions applied
to the baryonic TF relation improves the linearity of the correlation, increases the
slope and reduces the total scatter by 4\%.  

\end{abstract}

%\begin{keywords}
%techniques: photometric -- galaxies: star formation -- galaxies: stellar content
%\end{keywords}

\section{Introduction}

The stellar mass of a galaxy is one of its most fundamental
characteristics because it incorporates the endpoint of baryon mass evolution (from
atomic and molecular gas into stars and stellar remnants). Surprisingly, the path to
understanding dark matter is to first understand stellar populations in galaxies. For
example, the total baryonic mass of galaxies (gas and stars) tightly correlates with
the "flat" circular velocity, which is driven by the dark matter halo in the standard
cosmological context (McGaugh \etal 2000; Verheijen 2001; Lelli \etal 2019). The
properties of such baryonic Tully-Fisher relation (bTF) unavoidably depend on the
way we measure gas and stellar masses (e.g., Lelli \etal 2016). The gas component
has relatively small uncertainties of the order of 10\%: the gas mass is given by HI
observations of atomic hydrogen plus minor statistical corrections for heavier
elements and molecules (see McGaugh, Lelli, Schombert 2020). This leaves the stellar
mass as the remaining unknown to the total baryonic mass, where the stellar mass can
be deduced from the galaxy's luminosity with an assumed mass-to-light ratio
($\Upsilon_*$) deduced from stellar population models, or by kinematic determinations
of the mass surface density after subtracting the gas component (Martinsson \etal
2013).

The determination of stellar mass also has two components, one observational
(photometry), the other computational (stellar population models to deduce
$\Upsilon_*$).  Galaxy photometry, from the UV to the far-IR, has improved to the
point where uncertainties in galaxy luminosity are driven by the definition of which
galaxy components one wishes to examine rather than photometry errors
(see Stone \etal 2021).  While different galaxy types entail
different challenges in our photometry methods, the uncertainties are well-known and
less of a challenge to estimate.

The last remaining step is the application of $\Upsilon_*$ to the galaxy luminosity
and this involves several paths.  First is the specification of the correct
$\Upsilon_*$ to apply to a particular part of a galaxy, or to the galaxy as a whole.
Second, outlining the numerous details that go into the calculation of $\Upsilon_*$,
i.e. the star formation and chemical history of the portion of the galaxy to be
converted into a stellar population model and, then, into stellar mass
(Ge \etal 2021).  Third, isolating the inherent uncertainties due
the possible variations in the stellar tracks and chemical enrichment scenarios and
how they reflect into the uncertainties on the final stellar mass
(Lower \etal 2020).  The goal of this paper is to provide the
community with a straight-forward procedure to calculate the stellar mass of a galaxy
using spatial color information to guide the stellar population models and evolution
scenarios.  We divide galaxies into their bulge and disk components, although the
models are sufficient to calculate a total stellar mass simply from the galaxy's
morphological type plus total luminosity.

Throughout this paper, we test our stellar population models using
two galaxy samples with multiband photometry from the far ultraviolet (UV) to the
near infrared (IR). The SPARC sample consists of 175 spiral and dwarf galaxies with
high-quality HI rotation curves; the photometry is described in Lelli \etal (2016)
for the Spitzer 3.6 $\mu$m band and in Schombert \etal (2019) for the other bands.
The S$^4$G sample consists of 790 galaxies in common between the S$^4$G survey (Sheth
\etal 2010) and the SDSS survey (Blanton \etal 2017); the multiband photometry was
re-done by our team to measure accurate colors within the same physical aperture (see
Schombert \etal 2019) and to distinguish bulge and disk components (this paper).

\section{Stellar Population Models}

The core to any stellar mass project are SSP (single/simple stellar populations)
models produced by several groups over the past decade (see Conroy \& Gunn 2010).
SSPs are single burst models that start with a fiducial stellar distribution given by
an assumed initial mass function (IMF) at a set metallicity, then ages them with
standard stellar evolutionary tracks.  One can then use these SSPs to produce a
population of stars formed in a series of single burst events, where the length of
the burst is short enough to ignore the small spread in age.  During a short burst,
the chemical evolution is also negligible so the metallicity of all of the stars are
identical and unchanging per timestep.  Thus, a complicated star formation history
for a galaxy can be represented by a series of SSPs of varying ages and metallicity
to match an assumed star formation rate (SFR) as a function of time plus a chemical
evolution scenario.  At any particular timestep, the observables can be extracted,
such as a integrated spectra or color.

There are numerous variables that can enter into the construction of a SSP
(see Ge \etal 2019).  For example, one can introduce dust or
emission lines.  The IMF can be varied.  Or the evolution of AGB (asymptotic giant
branch) stars can be altered to simulate variations owing to metallicity changes.
The number of blue main sequence stars or blue stragglers can be varied to represent
star formation (SF) by strong cloud collision events that are richer in high mass
stars.  Turbulence can be introduced to increase the stellar rotation factors in
stellar evolutionary tracks.  These effects, and other nuances, were investigated in
Schombert, McGaugh \& Lelli (2019, hereafter SML) and make particular predictions to
the run of color vs color for galaxies, reflecting into our uncertainty in
$\Upsilon_*$ (see \S6). 

In addition, real galaxies do not form all of their stars in an instantaneous burst
like an SSP, even massive ellipticals that have initial star forming events that last
only over a few Gyrs (Thomas \etal 2005).  Thus, a star formation history (SFH) is
assumed, a distribution of star formation rate (SFR) with time.  The assumption of a
smooth SFR allows for the SFH to be broken down into a series of bins of unique age
that matches an individual SSP, and a metallicity per bin that can be varied to match
an assumed chemical evolution model.  Each bin in time is weighted by the number of
stars (formed during that timestep and normalized to the total mass of the galaxy),
and summed with all the older stars to produce a total color/spectra as a function of
time.

The goal of exploring the use of stellar population models is that, if some of the
characteristics of the chemical evolution and SFH of a galaxy is known, then one can
deduce a unique set of colors that, in turn, result in a unique value of $\Upsilon_*$
deduced from the models.  In other words, SFH plus metallicity maps into color that,
in turn, results into a unique model value of $\Upsilon_*$ with some, hopefully,
limited range of uncertainty.

An addition complication (and opportunity) is presented by the process of galaxy
formation where, frequently, a rotating galaxy is clearly composed of two distinct
stellar components, a bulge and disk (Sandage \& Tammann 1983).
Just based on their observational differences (bulges are redder and have spectral
signatures of older stars, Tasca \& White 2011), two different SFHs
and chemical enrichment schemes should be applied to each region resulting in
different $\Upsilon_*$ values.  These values are then applied separately to the
distinct luminosities that represent the populations in the disk and bulge.  Some
knowledge of the galaxy type and structure then allows those two components to be
added in a luminosity-weighted fashion to produce a total color and total
$\Upsilon_*$.  Our ultimate goal, then, is to present a simple empirical method of
relating galaxy color to $\Upsilon_*$, whether this be a disk or bulge or a user
defined combination of the two, with clearly defined uncertainties to that value.  

\section{Mapping the Star Formation History of Galaxies}

\begin{figure}
\centering
\includegraphics[scale=0.90,angle=0]{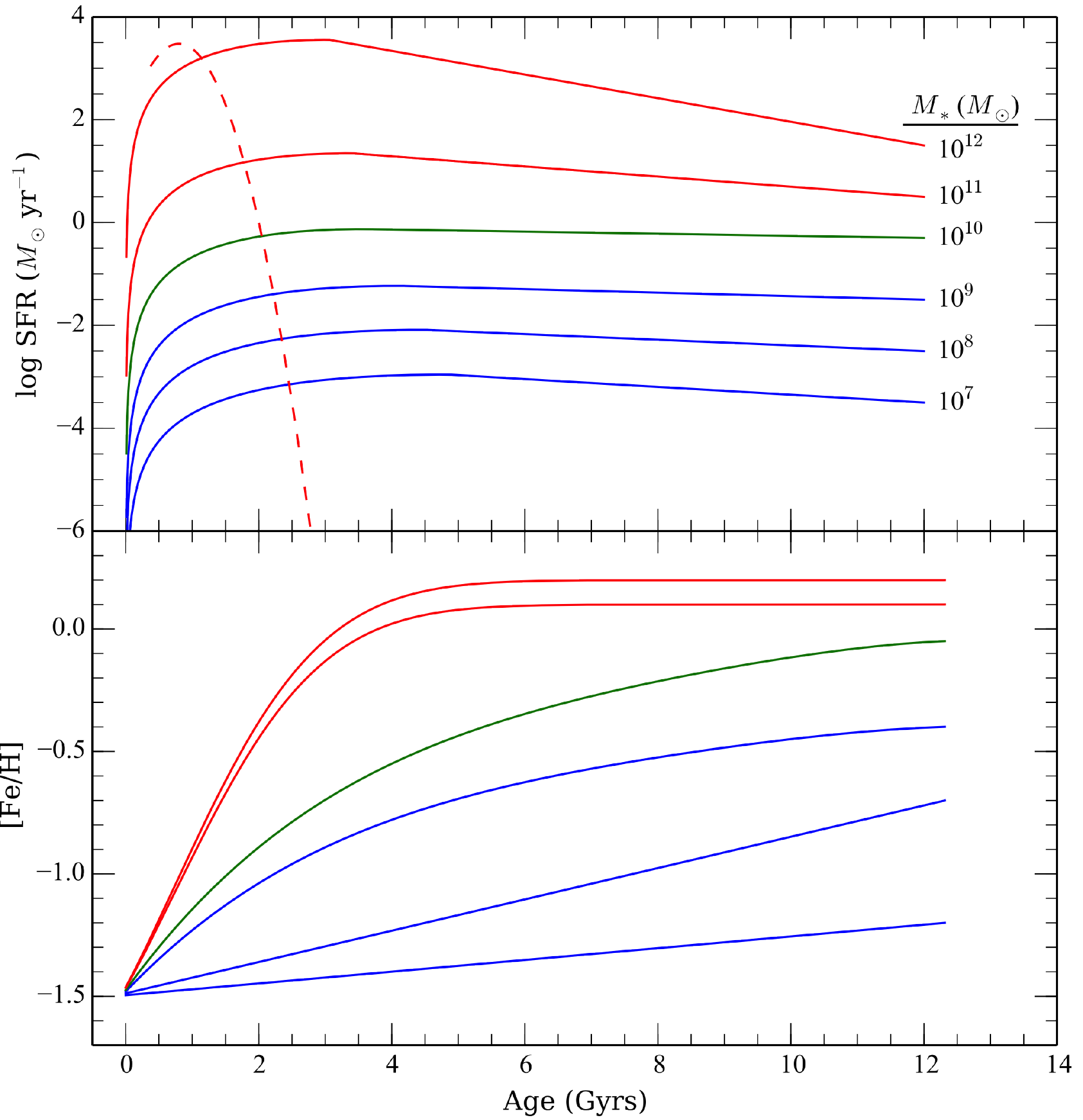}
\caption{\small Baseline star formation and chemical enrichment histories as a
function of stellar mass.  The red curves are for high-mass ($M_* > 10^{10}
M_{\sun}$) spiral galaxies that follow the Speagle \etal $z=0$ main-sequence.  The blue
curves are adjusted to match the low-mass main-sequence found by McGaugh, Schombert
\& Lelli (2017) with near constant SF over a Hubble time.  The green line is an
intermediate, canonical
SFH for a Milky Way sized system.   The early, strong SF for the high mass galaxies
results in the formation of red disks for early-type spirals.  The low intensity,
constant SF in low mass galaxies results in low density (i.e., LSB), low metallicity
stellar populations.  A 10 Gyrs burst model for a $10^{12}$ $M_{\sun}$ bulge is shown
as the red dashed line.  The bottom panel displays the adopted chemical enrichment
scenarios for the SFH in the upper panel.  The final metallicity is set by the
mass-metallicity relation for star-forming galaxies (Cresci \etal 2018).  Bulges are
assumed to have very fast enrichment to present-day values, thus, no chemical
evolution history is presented in the bottom panel.
}
\label{speagle_sfh}
\end{figure}

In general, we can divide the star formation histories in galaxies into two
simplistic models that map into the simple morphological division of early-type
versus late-type galaxies (see Peterkin \etal 2021).  The first
scenario is an early, strong burst with rapid chemical enrichment and a sharp decline
in the star formation after the burst (shown in Figure \ref{speagle_sfh}).  This
produces a present-day population that is primarily old and metal-rich, a common
feature of ellipticals, S0s and bulges.  A second scenario is given by a slowly
declining or constant SFR plus a steady chemical enrichment process proportional to
the SFR.  While simplistic, this scenario matches most of the characteristics of
early-type spirals, with red disks dominated by old stars, and late-type galaxies (Sc
to Irr), rich in young, low metallicity stars (see discussion in
\S4 of SML).  

The separation of bulges and disks, in terms of stellar populations, is similar to
the division between ellipticals and spirals.  The similarities between ellipticals,
S0s and bulges has its origins in the earliest papers on galaxy photometry (Sandage
\& Visvanathan 1978).  With the inclusion of S0s (as nearly non-star-forming disk
galaxies), one finds a subset of galaxies with old ages ($\tau > $10 Gyrs) and high
metallicities ($Z > 2Z_{\sun}$).  Bulges and ellipticals have similar structure,
being $r^{1/4}$ in shape, which is normally associated with the process of a rapid
conversion of gas into stars (MacArthur \etal 2008).  For our work, we will continue
to assume this simple version of the SFH for bulges, varying only the final
metallicities to explain the range in bulge color (Tonini \etal 2016, Calura \& Menci
2011).

One of the key diagnostics to unravel the SFH in star-forming disks is the so-called
main-sequence of star-forming galaxies, a correlation of stellar mass and current SFR
(see Speagle \etal 2014 and references therein).  As discussed in McGaugh, Schombert
\& Lelli (2017), the main-sequence of galaxies displays a distinct break at $M_* =
10^{10} M_{\sun}$ where higher mass spirals display the traditional exponentially
declining SFH after an early strong initial burst (Speagle \etal 2014).  This
explains their high stellar masses with a high current SFR, but red disk colors (a
great deal of past SF leaving a numerically larger old, red population in the disk).
Below $10^{10} M_{\sun}$ we find the realm of low mass disks and dwarfs that must
have nearly constant SF over a Hubble time to explain their current SFR and stellar
masses (McGaugh \etal 2017).  While there is some flexibility in the SFH of high mass
spirals, their colors rule out a number of more extreme SFHs (see discussion in SML).
Low mass disks have very little flexibility in their past SF, for even their current
SF barely creates enough stellar mass over a Hubble time and are fixed with a nearly
constant SF scenario with some room for an early epoch of burst SF so as to form
bulges or pseudo-bulges.  Their low metallicities from emission line measurements
(plus color-magnitude diagrams) also eliminates scenarios with strong bursts of SF
separated by long quiescent phases (Schombert \& McGaugh 2021).

The deduced baseline SFHs are shown in Figure \ref{speagle_sfh} for final stellar
masses of $10^7$ to $10^{12} M_{\sun}$ (see SML for a more detailed discussion).
While a majority of the stellar mass in disk galaxies is in the disk, the increase in
bulge mass can be seen reflected in the higher SFRs at early epochs for the high
mass spirals.  Other styles of SFH are considered, such as later initial formation
times, but, again, the more extreme scenarios can be rejected based on the comparison
between observed and predicted optical to near-IR colors.  These colors represent
various slices of time in the SFH (for example, UV colors represent stars of only a
few hundred Myrs) and measure any sharp changes from the current SFR over time scales
of Myrs to Gyrs (again, see SML for a fuller discussion, particularly
Figure 2 of that paper).

Likewise the chemical history of high and low mass disks are similarly constrained by
the correlation between stellar mass and current mean metallicity (Weldon \etal
2020).  The mass-metallicity relation, while having different slopes for disk and
bulges (Cresci \etal 2018, Li \etal 2018), still defines a monotonic increase in mean
galaxy metallicity with stellar mass.  This reflects the obvious correlation that
more SF results in more stellar mass while at the same time resulting in more element
recycling and, therefore, increasing chemical enrichment (see
Prantzos \& Boissier 2000).  The greatest unknown here is the rate of chemical
enrichment with time, which reflects into the ratio of metal-poor to metal-rich stars
per generation.

The strength of the initial SF epoch, combined with the known rapid chemical
enrichment phenomenon for strong bursts (Maiolino \& Mannucci 2019), results in a
proposed slow versus fast enrichment models (see bottom panel of Figure
\ref{speagle_sfh}).  Slow in the sense that low mass dwarf galaxies begin with
initial populations having mean [Fe/H] values between $-$2 and $-$1.5 but only reach
current values of $-$0.7 to $-$0.5 (in agreement with their very low SFRs averaged
over a Hubble time).  This process must proceed in proportion to the SFR, thus the
expectation of more metal-poor stars per generation than a galaxy with faster
enrichment (i.e., higher average SFR resulted in more stellar mass,
Gavazzi \etal 2002).  Over the same timescale, high mass disks
proceed from similar starting values to reach super-solar values in their
star-forming regions and the metal-rich bulge within a few Gyrs after initial SF
(i.e., fast).  Quantitatively, this results in more low metallicity old populations
(i.e., bluer) in low mass galaxies versus more metal-rich (i.e., redder) in high mass
disks for the same age.  This is the primary reason that LSB galaxies have bluer
colors compared to similar mass HSB galaxies (Schombert \& McGaugh 2015), although
the difference is negligible for this study.

Since past SFH defines the final stellar mass, which in turn determines the final
mean metallicity, the interrelated nature of the SF process allows the construction
of a fairly constrained set of galaxy color/spectra models from the deduced SFH and a
simple chemical enrichment model.  The sum of a finite set of SSPs of a specific age
and metallicity (at that age) produces a unique present-day galaxy spectra from which
we can extract characteristics, such as color and $\Upsilon_*$.  The adopted,
baseline evolutionary scheme is shown graphically in Figure
\ref{speagle_sfh and outlined in SML} where the final SFR, metallicity and stellar
mass are used to set the zeropoint for each model.  Low mass galaxies have current
SFR versus total stellar masses slighter richer than a pure constant SFR model, so
some early SF is assumed, which is in agreement HST imaging of nearby LSB dwarfs
(Schombert \& McGaugh 2021).  The presence of a significant $r^{1/4}$ bulge in higher
mass disks results in a model with a canonical strong initial burst followed by an
exponential decline as outlined by Speagle \etal (2014).  The primary consistency
check for these models is an accurate reproduction of the various two-color diagrams
(e.g, optical to near-IR colors as shown in Figure 6 of Schombert \& McGaugh 2014).

In our previous paper (SML), the models had difficulty reaching very red and very
blue colors.  On the red side, this is due to the problem that any recent SF
dominates the optical colors (pushing them blueward) even at very low levels typical
of Sa and S0 galaxies (Yildiz \etal 2017).  The reason may be threefold: (1) the
increasing contribution of a large bulge in early-type spirals, (2) increasing
reddening from the dust component (Schombert \etal 2013), and (3) possible shutdown
of the star-formation at least 1 Gyrs ago in Sa/S0s (Johnston \etal
2014).  On the blue side, the assumption of a constant SFR means that at very low
levels of SFR (log SFR $< -4$) luminosity from the star-forming component is very
weak (even in an LSB disk) and metallicity effects dominate the colors.  In fact, it
was impossible to produce $B-V$ colors less than 0.4 or $V-[3.6]$ colors less than
2.0 using these assumptions (extremely blue colors are primarily the domain of bright
starburst galaxies).

In order to extend the SF models to redder and bluer colors we have made two addition
assumptions.  On the red side, we assume that redder colors are primarily in higher
mass spirals with past histories of high SFR (building large and old disks).  Rather
than attempting to model the complex process of star formation quenching in red
disks, or an increased dust contribution, we have simply extrapolated models with
$V-[3.6]$ colors of 3.1 to 3.3 ($B-V$ from 0.8 to 0.9) in a linear fashion to cover
the reddest disks.  The resulting models produce red disk colors owing to an
increasing fraction of old, intermediate to high metallicity stars, rather than a
change in their SF histories.  With respect to integrated total colors, the reddest
spirals also have large bulges that begin to dominate their colors for type Sa/Sb
galaxies.  This does not seem to be an unreasonable extension to the models but we
indicate this extension with a dashed line in Figure \ref{avg_colors}.

The blue side is more problematic.  The bluest galaxies divide into two classes; 1)
LSB dwarfs with very low SFRs and very low metallicities (Schombert
\& McGaugh 2015) versus 2) starburst dwarfs, such as blue compact dwarfs (BCDs,
McQuinn \etal 2010, Gil de Paz \& Madore 2005) with high current SFRs relative to
their stellar mass.  The LSB dwarfs are nearly within our model colors for the lowest
current, and past, SFR.  But the models fail to reach $B-V$ colors less than 0.45,
nor $V-[3.6]$ colors bluer than 2.0, which describes about 10\% of the S$^4$G
survey (Sheth \etal 2010).  These bluer galaxies (both LSB and HSB
in nature) indicate our assumption of declining or nearly constant SF is violated in
the last 500 Myrs (the last slice of time that effects both optical and near-IR
colors).

To explore the low mass galaxies with bluer colors, we relax the constant SF
constraint for the last 500 Myrs and study three scenarios; 1) a decrease from the
past SFR by 50\%, 2) an increase from the past SFR by 10\% and 3) and increase from
the past SFR by 50\%.  These three scenarios were not chosen arbitrary but rather
were guided by our previous discovery of a variation of the low mass end of the
main-sequence with respect to far-UV colors (see SML and Cook \etal
2014).  As noted in SML, there is a distinct division on galaxies with blue versus
red $GALEX$ FUV-NUV colors (see Figure 1 of SML).  Galaxies with FUV-NUV colors less
than 0.25 lie above the constant SFR line, those with redder colors lie below.  This
indicates a slight change in the most recent SFR compared to the average SFR in the
past.  Numerical experiments focused on far-UV colors, which range from 0.0 to 0.6
for low mass galaxies in the Cook \etal 2014 and SPARC samples
(Schombert \& McGaugh 2014, see inset histogram of Figure
\ref{avg_colors}).  The decreasing SF scenario results in FUV-NUV colors around 0.5,
a weak increase produces colors near 0.3 and a strong burst results in FUV-NUV near
zero.  These models result in $V-[3.6]$ colors between 2.0 and 2.5 (depending on
final metallicity) and recover the bottom portion of the two color diagrams (see \S4
and Figure \ref{avg_colors}). 

We can also distinguish between very blue LSB dwarfs and starburst systems by
morphology and central surface brightness.  Typically, these very blue galaxies
correlate in morphology with their far-UV colors: the blue FUV-NUV galaxies (with
FUV-NUV $<$ 0.25) are high in surface brightness with Irr or IB type morphologies
(none earlier than Sc, Cook \etal 2014).  Whereas, redder near-UV
colors are found in LSB disks and dwarfs with blue near-IR colors from low
metallicities ($V-[3.6] < 2.0$), but not the sharp optical $B-V$ colors ($B-V < 0.4$)
seen in starburst BCDs.  We present both endpoints, with respect to $\Upsilon_*$, in
the following discussions.  In that respect, we find that most LSB dwarfs follow an
extrapolation of our SFH models to the lowest metallicities that have slightly rising
$\Upsilon_*$ values.  However, this is balanced by the fact that even a single O star
complex in a very low surface density environment drives $\Upsilon_*$ downward.  A
system with constant SF for 10 Gyrs, with a halt in SF for the last Gyr, has an
$\Upsilon_*$ of 0.5 at 3.6$\mu$m.  We use this as the bottom limit for our models,
but caution the reader that very blue starburst galaxies will have lower $\Upsilon_*$
by a factor of 2 to 3 (see \S4).

\section{Two Color Diagrams}

As a reality check, we require some observables other than those used to generate the
SFH model (i.e., SFR and mean [Fe/H]) to compare with model predictions.  Galaxy
colors can serve that role, particularly by comparison of widely separated colors in
wavelength space to capture the subtle effects from different types of stars; such as
blue main sequence (bMS), AGB and RGB stars.  The two colors of choice for this
comparison are $B-V$ and $V-[3.6]$ as they are influenced by very different ages in a
typical stellar population and display different sensitivities to metallicity
effects.  The optical color ($B-V$) primarily follows changes owing to recent SF (the
dominance of O to A stars).  The near-IR color ($V-[3.6]$) covers the region
dominated by RGB and AGB stars (the strength of the old population plus metallicity
from the position of the RGB).  There is a still a degree of degeneracy in using just
two colors (see Schombert \& McGaugh 2014 for a fuller discussion); however, for our
purposes of confirming a level of accuracy to our star formation models, this two
color diagram serves as a sufficient standard.  Those two colors are
shown in Figure 2 of SML for 790 galaxies in a combined S$^4$G and SPARC sample
selected for accurate ground-based and {\it Spitzer} photometry (see SML for details
of the extracted photometry).  

The photometry used are total colors derived from curves of growth to SDSS and {\it
Spitzer} images.  Thus, bulge and disk components are summed in this diagram,
although one can see the effect of an increasing red bulge component for early-type
spirals.  The curvature in this diagram signals a number of known stellar population
changes.  For example, galaxies with the bluest $V-[3.6]$ colors have the bluest
$B-V$ colors, but reach a plateau in $B-V$ around 0.45 for $V-[3.6] < 2.7$.  This is
due to the fact that $V-[3.6]$ quickly reddens with only a small increase in $B-V$
signaling the first onset of very red, yet young, AGB stars a few Gyrs after the
initial SF events.  At redder $B-V$ colors, the increasing bulge component drives the
$B-V$ values steadily to the red for early-type spirals, but the $V-[3.6]$ near-IR
color barely changes owing to a steady contribution from AGB stars and the fact that
metallicity changes very little for early-type disks (a flattening of the
mass-metallicity relation at high stellar masses).  The large spread in colors is
probably a indicator that this process is not as smooth as our baseline models assume
where the actual process of SF proceeds in a series of small bursts (a ``flicker"
SFH, see McQuinn \etal 2010).  These bursts then average in color space due to the
large timesteps needed to produce detectable changes in broadband colors.
Fortunately, the exact process of SF is irrelevant to the calculation of $\Upsilon_*$
as long as the process is monotonic on timescales of Gyrs.

\begin{figure}
\centering
\includegraphics[scale=0.75,angle=0]{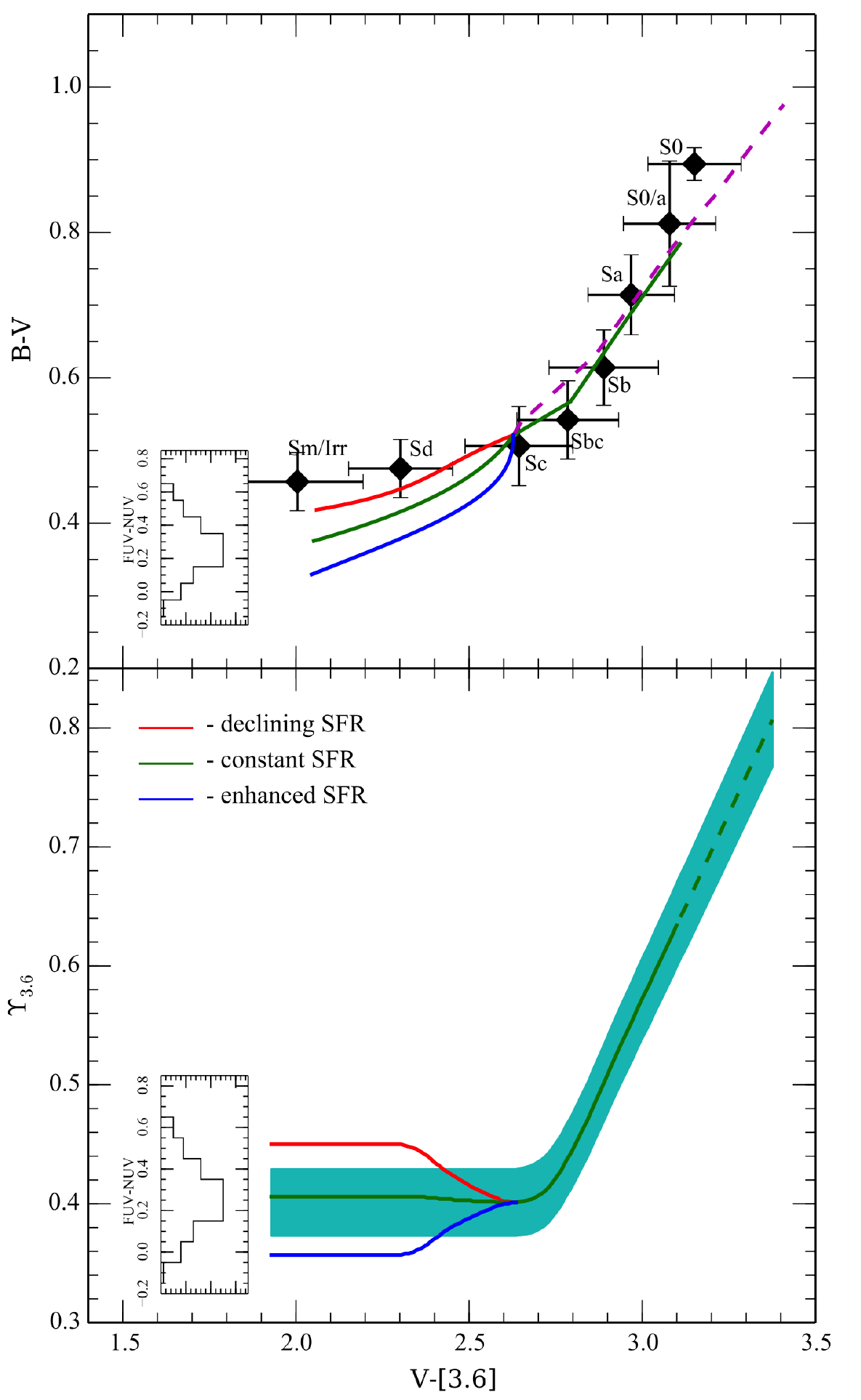}
\caption{\small The top panel displays the average optical to near-IR colors for the
SPARC and S$^4$G samples by galaxy type.  The errorbars display the 3$\sigma$
dispersion of the averages (not error in the colors).  The green track is the
baseline SFH model from SML (2019).  The dashed line is the new bulge+disk model that
captures the redder colors of early-type spirals (see \S5).  Three burst models are
shown for late-type galaxies; red for a declining SFR, green for a constant SFR and
blue for a weak increase in recent SFR.  A inset histogram of UV colors displays the
observed FUV-NUV colors matching the endpoint of each late-type model (where the mean
FUV-NUV color matches the constant SF model).  The bottom panel displays the effect
of those same models on the deduced mass-to-light ratios.  Here the solid lines
represent the original models adjusted on the blue side for late bursts of SF and 
extrapolated on the red side (dashed line) to capture older and more metal-rich red
disks.  Some knowledge of the
optical color or morphology of a blue galaxy can reduce the uncertainty in
$\Upsilon_*$ by 15\% on the blue end.  The cyan band indicates the range in
uncertainty in $\Upsilon_*$ owing to scatter in the mass-metallicity and main-sequence
relations.
}
\label{avg_colors}
\end{figure}

The division by galaxy type (RC3 T-type, Buta \etal 1994) is also evident in Figure 2
of SML, although the overlap in both colors is considerable.  Since bulge-to-disk
(B/D) ratio also correlates with galaxy type (Graham \& Worley 2008), we can use the
model colors outlined in \S3 to map each model onto a unique B/D to calculate the
relative weight of the bulge and disk components.  Operationally, we use the average
$B-V$ and $V-[3.6]$ colors for each galaxy T-type in the S$^4$G and SPARC samples
(using NED to extract the RC3 morphologies) and define a mean B/D per color.  The
average colors for each T-type are shown in Figure \ref{avg_colors} using their
standard Hubble designations.  It was found that galaxy T-types 1 and 2 had identical
colors, thus they are summed to form type 1.5 (basically all Sa galaxies).  Likewise,
types 6 and 7 were similar (late-type Sc) and were averaged to 6.5.  Everything later
than Sd were degenerate in their optical colors, so 8, 9 and 10 were summed into type
9 (basically a bin of all Sm/Irr types).  The errorbars are 3$\sigma$ dispersions,
not uncertainties on the colors.

The nominal baseline model colors from SML are shown as the green track in Figure
\ref{avg_colors} and does a fair job of matching the colors for galaxy types from Sa
to Sc.  The models do not reach the bluest near-IR colors for the galaxy types later
than Sd, which is probably due to deviations from the constant SF assumption in low
stellar density environments.  An extrapolation of the low metallicity models using
the scenarios discussed in \S3 results in the red and blue tracks in Figure
\ref{avg_colors}.  These two tracks simulate a slight rise or decline in the current
SFR that drive optical and near-IR colors on timescales less than 0.1 Gyrs.  This
type of behavior is also evident in CMDs for star-bursting dwarfs from Weisz \etal
(2011).

To extend the SFH models to redder colors, we need to deduce the effect of the bulge
colors on the total colors displayed in Figure \ref{avg_colors}.  The SPARC sample
provides a unique view of the interior color distribution because over one half the
sample has both {\it Spitzer} and SDSS imaging, plus another quarter of the sample
have $UBV$ values from the RC3.  Full imaging allows for the separation of the bulge
and disk components, and a direct comparison of the colors of those components can
guide the construction of our bulge+disk color models.  The SPARC sample has a full
range of galaxy types, central surface brightness, morphology and luminosities so
selection effects are minimized.

For our analysis, we define a break point in the 3.6 surface brightness profile where
the bulge dominates over the disk.  Interior to that radius, an elliptical aperture
(defined again by the 3.6 isophotes) is applied to each SDSS $g$ and {\it Spitzer}
3.6 frame.  These luminosities are subtracted to deduce a bulge color.  The
corresponding disk color is deduced by subtracting the bulge luminosity from the
total luminosity of the galaxy (based on curves of growth).  Experiments with varying
break radii finds that the bulge and disk colors are stable to the 2\% level,
typically owing to the fact that bulge surface brightnesses are much higher than disk
surface brightnesses, but the bulge area is smaller than the disk area.

%\begin{figure}
%\centering
%\includegraphics[width=\columnwidth,scale=0.80,angle=0]{bulge_disk.pdf}
%\caption{\small Bulge versus Disk colors for the spiral galaxies in the SPARC sample.
%Red symbols are classic $r^{1/4}$ bulges in early-type spirals, blue symbols are
%pseudo-bulges found in late-type spirals.  The correlation between disk and bulge
%near-IR colors is clear, but weak.  An average difference of 0.3 mags between bulge
%and disk colors is indicated with the dashed line, and assumed in our model
%constructions using the disk color to define both the suitable SFH model and the B/D
%ratio.  The bulge SSP and disk SFH model are then iterated to converge on the value
%of the B/D that corrects predicts the total color.
%}
%\label{bulge_disk}
%\end{figure}

There are solid correlations between bulge/disk colors and the total luminosity of a
galaxy such that brighter spirals have redder colors (Kennedy \etal
2016).  These are mostly dominated by the correlation between stellar mass and
metallicity (where luminosity is a proxy for stellar mass and color is a proxy for
mean metallicity), but is somewhat surprising as there is no particular dominant
scenario where bulge and disk formation are synchronised (although, clearly, larger
initial gas mass will lead to larger bulges and larger disks which drives the final
color).  In addition, as the main-sequence relation indicates that all spirals below
$10^{10} M_{\sun}$ have nearly constant SF for a Hubble time, they all also have the
roughly the same distribution of stellar ages in their stellar population
(Weisz \etal 2011).  Therefore, the correlation with color appears
to be driven by chemical enrichment rather than age although a sharper decline in SF
for high mass spirals will result in more older (and redder) stars.

The difference between the bulge-luminosity and disk-luminosity correlations are
notable.  As expected, bulges are redder than disks at any particular bulge
luminosity; however, the difference in colors is roughly constant with disk color.
Bulges are, on average, 0.3 mags redder than disks in $V-[3.6]$, although on a
case-by-case basis there is a great deal of variation.  The distinction is sharper
for classic r$^{1/4}$ shaped bulges versus disks; decreasing in difference for
pseudobulges whose low luminosities distort the separation of bulge and disk light.

We can use this fact to constrain the disk+bulge models in \S5, where for a given
total color, there is a unique disk and bulge color combination that matches the
expected B/D ratio of that total color (since color correlates with morphology).  In
other words, for each disk model that produces a disk color $X$, there is a bulge
model of color $X+0.3$ that serves as the bulge population color.  We can consider
the scatter in the bulge/disk color relation as the uncertainty in B/D ratios for
purpose of evaluating a particular model.

\section{Bulge+Disk Models}

The SFH models outlined in the previous section are best applied to colors in regions
of a galaxy with common SF and chemical enrichment scenarios.  There is strong
observational evidence that bulge and disk components have very different kinematic
and SF histories (van den Bosch 1998).  In particular, the different structural
shapes plus opposite optical to near-IR colors signals different evolutionary paths.
The most significant difference is, of course, the distinct lack of current SF in
bulges compared to the fact that ongoing SF dominates the appearance of disks.  Thus,
it seems inappropriate to use the same models for star-forming disks to deduce
$\Upsilon_*$ in bulge regions.

The best scenario to model bulges is the well-established single burst model used so
successfully to predict the colors of ellipticals (see Samland \& Gerhard 2003).  Under
this approximation, the bulges are assumed to be single burst population of singular
age with a metallicity that is proportional to their total mass (see Schombert \&
Rakos 2009).  For our study, we have adopted an age of 10 Gyrs and a range of final
metallicities from [Fe/H] = $-$1.0 to $+$0.2 (see Costantin \etal 2021).  This
metallicity distribution covers the range in optical and near-IR colors displayed by
bulges in the SPARC sample on the assumption that metallicity is the primary driver
of bulge color.

\begin{figure}
\centering
\includegraphics[width=\columnwidth,scale=0.80,angle=0]{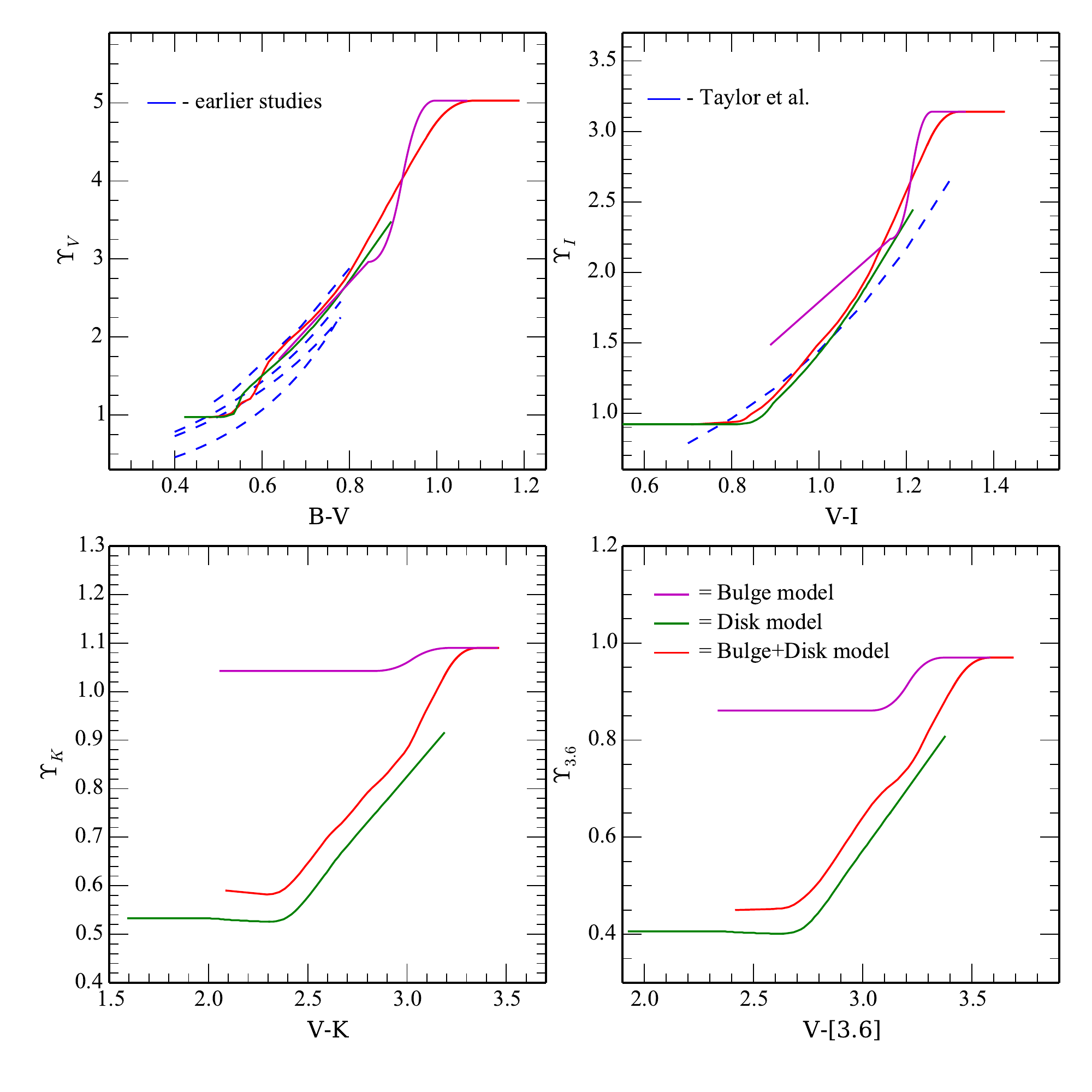}
\caption{\small Four $\Upsilon_*$ models for the optical and near-IR ($V$, $I$, $K$
and IRAC 3.6).  The green lines are the SFH models from SML (extended to redder
and bluer colors as discussed in text). 
A 10 Gyr bulge model is shown as a purple line in each panel.  The merged bulge+disk
models are shown in red.  As the models progress from the near-IR to the optical, the
difference between the various models merge as the youngest stars dominate in the
optical.  The slope of the color-$\Upsilon_*$ relation steepens with bluer filter,
meaning that errors in optical colors will reflect into larger $\Upsilon_*$ errors
compared to the near-IR.  Also shown in the $B-V$ panel are previous
color-$\Upsilon_*$ studies discussed in Schombert \& McGaugh (2014,
blue dashed lines) and in the $V-I$
panel the relationship defined by Taylor \etal (2011) for SDSS $i$ photometry.
Bulge+disk models are needed to extend these older studies to redder colors, but
agrees well with our newer models.  The Taylor \etal relation is in agreement with 
intermediate colors, but deviates significantly for red, early-type galaxies.
}
\label{bd_side}
\end{figure}

While there is not a perfect correlation between bulge and disk color, we can use the
trend in disk to bulge color to select a bulge color (and, thus, a unique model)
based on the disk color and model.  With our baseline scenario, we assume a mean
difference of 0.3 mags in $V-[3.6]$ color from the disk to the bulge.  Thus, when we
blend the disk and bulge components (for an assumed B/D ratio), we can use a bulge
model that matches the disk color (increased by 0.3 mags).  In general, these bulge
colors represent a slightly higher metallicity than the disk model, but one would
expect the bulge metallicity to be slightly higher than the final disk metallicity
due to a more rapid onset of chemical enrichment from the bulge's initial burst of SF
as well as expected reservoirs of low metallicity gas available to disk.

Lastly, there is the assignment of the B/D ratio for the summed bulge and disk
components.  The B/D ratio is primarily a function of galaxy type (although,
operationally, it is galaxy type that is dependent on the visual B/D for types 
Sa to Sc).  Galaxy type is also strongly correlated with galaxy color (Graham \&
Worley 2008).  Therefore, we can use model color to assign a galaxy type and, thereby,
a B/D ratio.  A small amount of iteration is required, as we start with a pure disk
color, then extract a mean B/D ratio followed by the application of the luminosity
ratio using a bulge color from the bulge versus disk color diagram.

To summarize, models of star-forming disks are produced using a grid of population
models of ages and metallicities given by the mean SFHs taken from the main-sequence
relationship (normalized by stellar mass, outlined in SML) and the mass-metallicity
relation, as shown in Figure \ref{speagle_sfh}.  The near-IR disk colors are calculated
from the model, then used to assign a galaxy type by color (near-IR colors are used as
they are less sensitive to sharp changes in the recent SFR).  From the galaxy T-type,
a B/D is assigned based on the grid of B/D and morphological type (Graham \& Worley
2008).  A 10 Gyrs burst bulge model is calculated and summed with the disk colors to
deduce a total near-IR color.  This value will typically underestimate the correct
B/D as the disk colors will always be bluer than the bulge colors for the first
estimate.  A short iteration is made of the disk and bulge models to match the total
$V-[3.6]$ color to the correct B/D ratio, then the resulting model is converted into
colors from optical to near-IR.  Note, those same models have unique $\Upsilon_*$
values for the disk and bulge components, which are summed (weighted by near-IR
luminosity) for a total $\Upsilon_*$ for each model color.

The track of total color ($B-V$ versus $V-[3.6]$) for these bulge+disk models is
shown in Figure \ref{avg_colors}.  At the bluest colors, the contribution from the
bulge is negligible and the colors converge to the baseline disk SFH models.  As the
bulge light increases in fractional contribution, we see the colors moving redward
compared to the baseline models.  Again, the sharp rise in near-IR color ($V-[3.6]$)
for the bluest galaxies signals the first epoch of AGB stars entering the near-IR
bandpasses.  Both the baseline and bulge+disk models are slightly bluer then the
average optical color of late-type galaxies, perhaps due to an underestimate of the
AGB component (Schombert \& McGaugh 2015).  As the models approach the reddest
galaxies, the bulge contribution dominates and approaches the red region representing
the colors of ellipticals (Schombert 2016).  Solar metallicity models match the
reddest spirals, but the models overestimate the near-IR colors for non-star-forming
E/S0s, again probably due to mismatch of the AGB contribution for old populations
plus an underestimate of the contribution from a hot component such as old, low
metallicity, blue horizontal branch (BHB) stars (see Schombert \& Rakos 2009).  An
increase in mean metallicity by 0.2 dex on the AGB component brings the Sb to Sc
values in agreement with the colors.  The effect of these changes on $\Upsilon_*$ is
discussed in \S6; however, in general, the predicted colors are in good agreement
with the observed colors of the S$^4$G and SPARC samples.

Matching the blue disk models to the bulge+disk models (blueward of $V-[3.6]$ = 2.6)
requires an estimate of the style of SF in the last 500 Myrs.  Our three scenarios,
matched to far-UV colors, are shown in the top panel of Figure \ref{avg_colors} for
the two color diagram and in the bottom panel for their effect on the correlation
between color and $\Upsilon_*$.  Our original declining SF model predicts a slight
rise in $\Upsilon_*$ below $V-[3.6]=2.6$.  However, the constant SF model predicts a
constant $\Upsilon_*$ below $V-[3.6]=2.6$ and the increasing SF model predicts a
slightly lower $\Upsilon_*$.  The range in $\Upsilon_*$ at 3.6$\mu$m is from 0.35 to
0.45.  A mean of 0.5 has been used in past studies (see Lelli \etal 2016).  But
better UV or near-blue color information can reduce the uncertainties by 15\% (we
note that FUV-NUV or $B-V$ or a morphology estimate are all equally effective at
defining $\Upsilon_*$ on the blue side).

The ultimate goal of the SFH models is, of course, the extraction of a $\Upsilon_*$
for each model (i.e., each galaxy color).  These values are filter dependent, and we
express their values as a function of the color $V$-$X$, where $X$ is the filter of
interest.  Four examples are shown in Figure \ref{bd_side} for filters $B$, $I$, $K$
and IRAC 3.6.  The blue curves are the same as those from SML, extrapolated now to
the cover the reddest and bluest galaxies in the SPARC dataset.  The bulge model is
constant in $\Upsilon_*$ in the near-IR until metallicities rise high enough for a
decrease in the AGB population luminosity and sequential rise in $\Upsilon_*$.  The
bulge+disk model have slightly higher $\Upsilon_*$ values at all points, representing
the higher $\Upsilon_*$ from the bulge population, and merge with the disk and bulge
models at the bluest and reddest colors.  The extrapolated disk model is for LSB
dwarfs on the main-sequence diagram and with very low metallicities using the
constant SF model from Figure \ref{avg_colors}.  

\begin{figure*}
\centering
\includegraphics[width=\columnwidth,scale=0.90,angle=0]{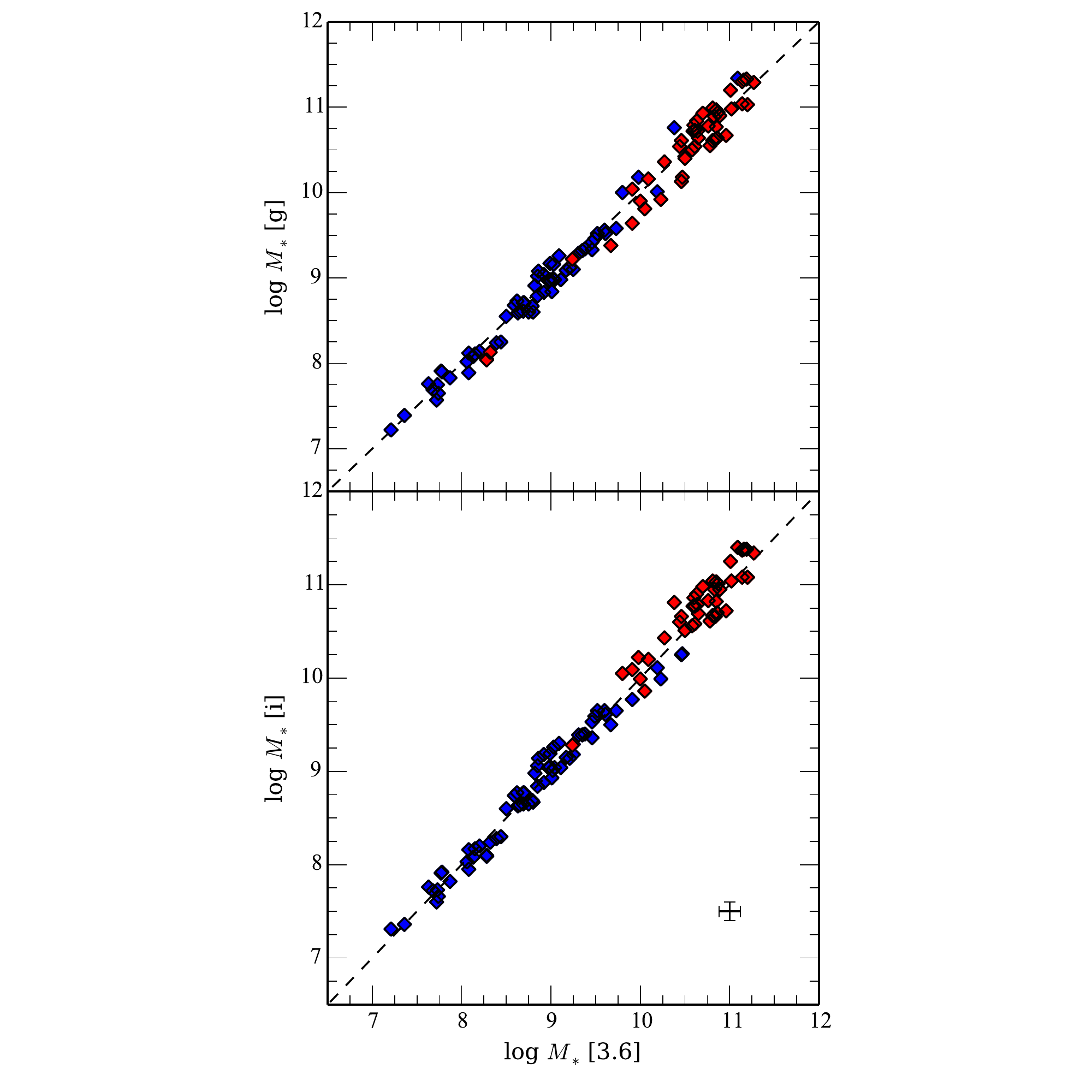}
\caption{\small A comparison of deduced stellar masses using SDSS $g$, $i$ and {\it
Spitzer} 3.6 photometry plus the new bulge+disk models from Figure \ref{bd_side}.
The SPARC sample is divided into red and blue galaxies based on colors greater or
less than $g-i = 1.05$ or $V-[3.6] = 2.9$.  The agreement across various optical and
near-IR filters is excellent with no obvious systematics.  The dispersion increases
slightly to bluer filters in expectation with the increase in color versus
$\Upsilon_*$ slopes for bluer colors.
}
\label{multi_masses}
\end{figure*}

In general, one finds that $\Upsilon_*$ increases steadily with redder colors (a
higher percentage of older stars with high $\Upsilon_*$ values).  Also, the slope
of the color versus $\Upsilon_*$ relation becomes shallower with increasing
wavelength (and the dynamic range is reduced).  Aside from lower extinction effects
owing to dust, this is the primary reason that near-IR photometry produces superior
$\Upsilon_*$ estimates.  As one goes to optical filters, small errors in galaxy color
produce larger errors in $\Upsilon_*$ compared to near-IR filters.  This same trend
is found by numerous previous studies.  For example, also shown in the $B-V$ panel of
Figure \ref{bd_side} are the results from four previous studies on $\Upsilon_*$ in
the optical (Portinari \etal 2004, Zibetti \etal 2009, Into \& Portinari 2013, and
Roediger \& Courteau 2015, see also SML).  Although they all use varying assumptions
on the SFH of star-forming galaxies, the slopes are remarkable similar and the mean,
at any particular $B-V$ value, is similar to both our SF and bulge+disk models.  A
webtool to compute $\Upsilon_*$ is available for the
community\footnote{http://abyss.uoregon.edu/$\sim$js/sfh}.

A different technique was presented by Taylor \etal (2011) using a sophisticated
Bayesian parameter fitting to the optical SED from SDSS photometry.  They present a
SDSS $g-i$ vs $\Upsilon_*$ relation and an analytic fit arguing that the use of UV
and near-IR colors only marginally improves the accuracy of estimating $\Upsilon_*$
from SDSS $i$ photometry.  Their $g-i$ relationship is shown in the $V-I$ panel
(green dashed line) and is an excellent match for intermediate colors, but
underestimates $\Upsilon_*$ for the redder colors compared to our bulge+disk model.
This difference is primarily due to our empirical treatment of the AGB contribution,
particularly in near-IR filters.  While their analysis of the deficiencies in the
near-IR differs from our conclusions, we note that the dynamic range in the {\it
Spitzer} bandpasses is a factor of six less than in the SDSS $i$ filter, with a
corresponding decrease in uncertainty.  In addition, their $g-i$ estimator severely
underestimates the $\Upsilon_*$ values for early-type spirals and ellipticals, which
can have a significant impact on the results from lensing studies that late-type
and early-type galaxies differ in their positions on the radial acceleration relation
(see Brouwer \etal 2021).

As a consistency test, we compare the stellar masses for the SPARC sample deduced
from the bulge+disk models using SDSS $g$, SDSS $i$ and {\it Spitzer} IRAC 3.6 images
in Figure \ref{multi_masses}.  The only sources of error in this comparison are the
errors in the photometry (both absolute and in color) and the uncertainties in the
parameters that went into the bulge+disk models.  While there is a wide range of
metallicities and ages that correspond to a particular color (and, thus, a value for
$\Upsilon_*$), the main-sequence and mass-metallicity relations appear to constrain
the range in models enough to produce an excellent agreement between the deduced
$\Upsilon_*$ values from the near-blue to the near-IR.  There are no obvious
systematics by mass or galaxy color.  The dispersion is higher than the expected
error based solely on the photometry, this would indicate that the scatter in color
reflects real scatter in the age and metallicity of the underlying stellar population
and the use of the $\Upsilon_*$ models is limited by knowledge of the stellar
characteristics beyond the mass-metallicity and main-sequence relations.

\section{Uncertainties in Deducing $\Upsilon_*$}

The key to our analysis of $\Upsilon_*$ is the fact that a unique SFH and metallicity
distribution for a composite stellar population maps into a unique SED (i.e., color)
and, therefore, a unique $\Upsilon_*$.  The uncertainty in $\Upsilon_*$, then, has
three components; 1) errors in the galaxy photometry, 2) errors in the SSP SEDs and
3) variation in the main-sequence and mass-metallicity relations (as they reflect
into the deduced stellar population model).  First, we consider errors in the galaxy
photometry.  These arises not in an error of the total luminosity (which is to be
converted into a total stellar mass), but rather into an error in the color used to
deduce $\Upsilon_*$ from the bulge+disk models.  An error in color magnifies as one
goes to bluer filters, for the slope of the color versus $\Upsilon_*$ curve steepens.
A typical magnitude error of 0.01 in luminosity, and 0.02 in color, corresponds to an
uncertainty of 10\% in $\Upsilon_*$ at $V$, but only 4\% at 3.6$\mu$m.  This is a
strong argument for continuing to use near-IR bandpasses to measure stellar mass as
the shallower $\Upsilon_*$-color slope minimizes the effect of photometric error.  We
constrain the models using the mean colors per galaxy type, but assign the
$\Upsilon_*$ values based on the individual galaxy colors that, in the end, results
in an near equal contribution to the error budget.  The errors in the galaxy
photometry are well-known and discussed in the various photometry papers (see SML).
The mean error varies slightly with galaxy type, but an error of 0.05 describes the
entire sample within 20\%. 

The errors in the actual construction of the SSPs is quite small as the newest
stellar libraries are highly detailed.  The largest uncertainty in a composite SSP is
how independent exotic components, such AGB or BHB stars, are included.  While the
main sequence and RGB stars are fairly well-defined by the metallicity and age of the
stellar population, the contribution from AGB stars can vary between various studies
(see Eftekhari, Vazdekis \& La Barbera 2021).  We are guided, again, by the two color
diagrams to constrain the more extreme models discussed in SML.  While models with
enhanced blue main sequence stars or suppressed AGB populations can explain the edges
of the two color distribution of galaxies, they are frequently inconsistent with the
mean colors of each galaxy type (see Figure \ref{avg_colors}).  And the dispersion in
color, per galaxy type bin, is consistent with the dispersion solely in metallicity
and star formation history (see below).  Thus, we adopt a typical error in the
photometry as it maps into a range of the baseline models.  For the SPARC and S$^4$G
samples, the mean error in $V-[3.6]$ color is 0.05, that resolves into a model
uncertainty in the near-IR of $\Delta\Upsilon_*$ = 0.04.

Lastly, are the uncertainties introduced by a range in SFR and metallicity
enrichment.  An estimate of their effect on the models can be obtained by considering
the scatter in the main-sequence and mass-metallicity relations as they reflect into
the models.  The main-sequence relation has two legs, the high mass end with a
relatively shallow slope and the low mass end with a slope of near unity implying
constant SF over a Hubble time.  The SFH of low mass galaxies is highly constrained
due to the limited time to produce their current stellar masses at their current
SFRs.  The UV color correlation within the main-sequence (Schombert, McGaugh \& Lelli
2019) plus numerous resolved CMDs for nearby dwarfs, indicates that a series of
micro-bursts (i.e., SF flickering, see McQuinn \etal 2010) rather than strictly
constant SF will satisfy the lower main-sequence; however, each micro-burst is
sufficiently low in intensity that it produces a composite stellar population that
differs very little in mean colors from a constant SF model (except for the bluest UV
colors).  The high mass end of the main sequence contains galaxies with much higher
current SFRs, but variations from the Speagle \etal models result, primarily, in
early production of bulge stars and the resulting Hubble sequence of early-type
spirals.  Numerical experiments using the observed scatter in the main-sequence,
mapped into models with varying SFR values, displays a dispersion in $V-[3.6]$ color
of 0.04 on the blue end to 0.06 on the red side.  This results in a dispersion of
0.02 to 0.06 in $\Upsilon_*$ for SFH effects.

The uncertainty in metallicity is deduced from the scatter in the mass-metallicity
relation (see Kewley \& Ellison 2008).  Although slightly higher at lower
metallicities, due to the low number of dwarfs in their sample, the mean dispersion
is approximately 0.12 dex in [Fe/H] for the range in stellar masses considered by our
study.  Again, with numerical experimentation, this maps into a range of $\Upsilon_*$
between 0.02 and 0.05.  Thus, the resulting uncertainty from the SFH and metallicity
model inputs, if added in quadrature, are approximately 0.04 on the blue, low mass
end and 0.06 on the red, high mass end.  We note that this also maps into dispersions
in color of 0.15 to 0.25 for $V-[3.6]$ which nicely brackets the observed dispersion
in colors for the SPARC and S$^4$G samples.  We conclude that most of the dispersion
in galaxy color is, then, due to variations in SFH and chemical enrichment with
respect to our baseline models.  The dominant source of error in $\Upsilon_*$ is the
range in possible SFH and metallicity that matches a particular galaxy color.

In summary, the uncertainty in $\Upsilon_*$ from photometric errors varies
considerable with wavelength being minimal in the near-IR, although Figure
\ref{multi_masses} demonstrates that optical filters can achieve the same level of
accuracy as near-IR filters with good colors.  For highly accurate galaxy
luminosities, the current limit in $\Delta\Upsilon_*$ has its origin in the
dispersion in possible SFHs and internal metallicity distributions.  This value is
approximately 0.06 at 3.6$\mu$m (0.05 dex in log $M_*$).  The uncertainty increases
slightly on the blue and red ends of the color versus $\Upsilon_*$ relationship due
to sharp, recent star formation events on the blue end, and the effects of dust and
metallicity on the red end.  This is shown, graphically, in Figure \ref{avg_colors}
as the shaded band around the star-forming disk model.

\section{Conversion from Luminosity to Stellar Mass}

Armed with these new models there are now four methods to convert galaxy luminosities
into stellar masses by photometric means.  We will demonstrate that
our technique using $\Upsilon_*$ values deduced from {\it Spitzer} 3.6$\mu$m images
for these luminosities has the narrowest range in $\Upsilon_*$ and the smallest
scatter.  The four methods are as follows; first, the application of a mean
$\Upsilon_*$ value to the bulge and disk luminosities.  This was the technique used
in the earliest SPARC papers (Lelli \etal 2016) where an $\Upsilon_*$ of 0.5 was
assumed for disks (the mean value from the pure disk models) and a value of 0.7 was
assumed for the bulge component.  The second method is to deduce a more accurate
$\Upsilon_*$ value using the color information of each component.  Thus, a
$\Upsilon_*$ value for the disk is taken from the disk model and the color of the
disk region.  Likewise, a bulge $\Upsilon_*$ is deduced from the bulge color.  These
two values are then used to the observed disk and bulge luminosities and summed.  The
third method is used if only a total aperture color is available.  Then, the total
$\Upsilon_*$ is deduced from the total color and the bulge+disk models in Figure
\ref{bd_side}.  This method will be the most useful for large galaxy surveys where
the spatial information is not saved and only total colors and luminosities are
extracted from the datasets.  Lastly, if there is no color information, only a
3.6$\mu$m flux, then one can estimate the total $\Upsilon_*$ value from the galaxy
morphology.  Since morphology is correlated with color, one can go from estimated
color to $\Upsilon_*$ with the bulge+disk model.  There is an expectation that the
second method is the most accurate as it uses all of the color information of a
galaxy in a spatial way, with the third method producing similar values,
statistically.  The last method, by galaxy type, should be the least reliable and we
can compare all our techniques to the original prescription from Lelli \etal (2016).

\begin{figure}
\centering
\includegraphics[width=\columnwidth,scale=0.80,angle=0]{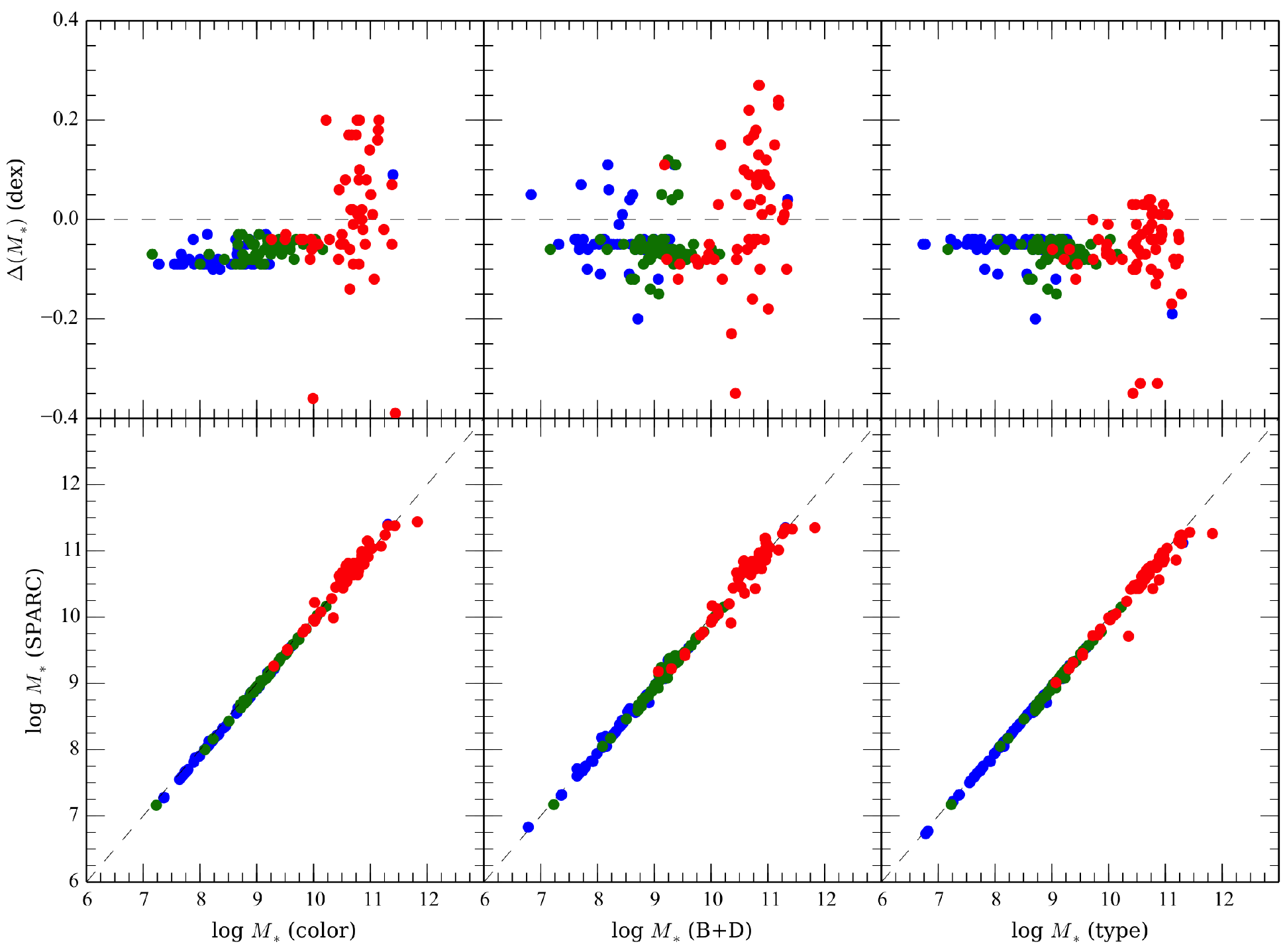}
\caption{\small A comparison of the original $\Upsilon_*$ prescription (SPARC, Lelli
\etal 2016) 
on the y-axis versus the three new color methods.  The first panel uses the color of the
separate disk and bulge components to derive $\Upsilon_*$ (using 3.6 luminosities).  The
second panel uses the combined B+D model and total galaxy color.  The third panel
uses the galaxy type to predict the total color, then applies the B+D model.  The upper
panels display the difference from unity in log space.  Note the disk model assumes a
new SFH of constant SFH for very blue galaxies (see Figure 2) with a mean
$\Upsilon_*$ of 0.41 versus the assumed value of 0.5 in our previous papers.  Red symbols galaxies with classic,
$r^{1/4}$ bulges, green for pseudo-bulges and blue for pure disk systems.  The most
significant difference is the higher stellar masses for early-type spirals with the
correct inclusion of bulges.
}
\label{mass_compare}
\end{figure}

To test the different methods, we have plotted the calculated stellar masses for the
SPARC dataset in Figure \ref{mass_compare}.  Of the 175 galaxies in the SPARC
dataset, 132 have SDSS imaging available from the DR16 archive.  Another 22 have
Johnson $V$ aperture values available from NED.  Of the 132 with SDSS imaging, 50
have classic $r^{1/4}$ bulges, 41 have pseudo-bulges (defined as a central
concentration of light, but without a power-law profile) and 41 are pure disk systems
based on examination of their surface brightness profiles (each type
displayed as a different symbol color in Figure \ref{mass_compare}).  

As can be seen in Figure \ref{mass_compare}, the late-type galaxies display very
little change between the four methods.  The primary offset is due to the change in
the baseline disk model from a declining SFH to a constant SFH, which results in a
shift from 0.50 used in the original SPARC sample, to a mean value of 0.41.  Other
than this constant shift, the disk model varies very little with color between
$V-[3.6] = 1.5$ and 3.0 (where 90\% of the disk colors lie).  The early-type spirals
display more scatter, again due to the nature of a color relationship where blue
bulges have lower $\Upsilon_*$ values from the canonical value of 0.7 and red bulges
have higher values.  It is worth noting that using galaxy type as a proxy for color
recovers a great deal of the estimated stellar mass compared to the canonical values,
but consistently over estimates the mass value compared to color models.

An independent check on stellar mass-to-light ratios is provided by
the velocity dispersions of face-on galaxies (e.g, Bershady \etal 2010, Martinsson \etal
2013). These appear to be in tension with stellar population models (Angus \etal
2016), but this seems to be caused by a difference in the populations tracing the
stellar mass and those dominating the velocity ellipsoid traced by the available
spectral lines (Aniyan \etal 2021). For the case of NGC 6946, our photometric
estimate of the stellar surface density is in excellent agreement with that inferred
kinematically by Aniyan \etal (2021).

\section{Effects of the bTF Relationship}

\begin{figure}
\centering
\includegraphics[width=\columnwidth,scale=0.80,angle=0]{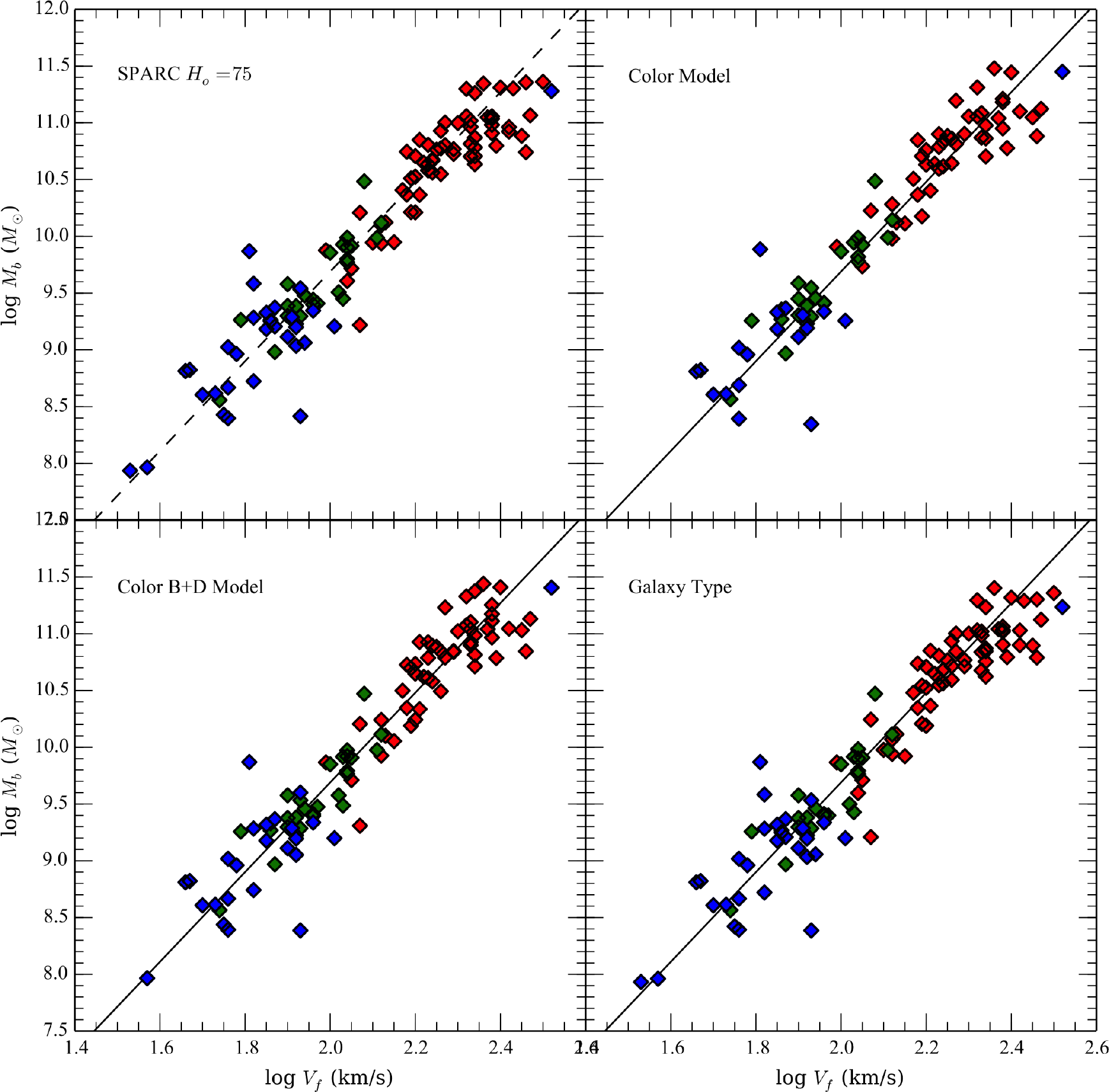}
\caption{\small The changes in the baryonic Tully-Fisher relation using different
$\Upsilon_*$ models.  The upper left is the original SPARC dataset using EDD
distances and flow models with $H_o=75$.  The upper right is the same kinematic data and distances,
but calculating $\Upsilon_*$ from the colors of the bulge and disk components.
Galaxies without spatial color information were left out of the sample.  Bottom left
is the two color B+D model and bottom right using only the galaxy type to set the
galaxy color.  Symbol colors are the same as Figure \ref{mass_compare} for bulges,
pseudo-bulges and pure disks.  The bTF is significantly more linear with either color
model, but scatter around the C/TRGB fit is only reduced by 5\%.
}
\label{btf}
\end{figure}

The primary impact of new $\Upsilon_*$ models is their effect on the deduced stellar
masses for baryon mass to kinematic relationships, such as the baryonic Tully-Fisher
relation (bTF).  The effect of the new models on the bTF can be seen in Figure
\ref{btf}.  The original bTF for the SPARC sample uses distances from the EDD
database (Tully \etal 2016) and velocities from Lelli \etal (2019).  A full
description of the dataset can be found at Schombert, McGaugh \& Lelli (2020).  The
original bTF, used an $\Upsilon_*$ of 0.5 for disks and 0.7 for bulges and is shown
in the upper left panel.  The upper right displays the same relation using the actual
color information of the disk and bulge.  The lower left panel displays the
application of B+D model to the total color of the galaxy.  And, lastly, the bottom
right panel uses the relationship between galaxy morphological type and color to
deduce $\Upsilon_*$ from the B+D model.

There are two points to note.  First, the downward trend at high baryonic masses is
significantly reduced with the color models where early-type spirals have higher
stellar masses due to more accurate $\Upsilon_*$ from their disk and bulge colors.
As high baryonic mass galaxies typically have low gas fractions, this change in
$\Upsilon_*$ is more critical to their final baryonic mass values than for the low
mass, high gas fraction galaxies.  This upward adjustment of early-type spirals
(basically due to the inclusion of an accurate bulge component) in the bTF supports
the observation that the bTF is surprisingly linear in log space with a power-law
slope of 4, in contrast to predictions from a $\Lambda$CDM cosmology (see also Di
Teodoro \etal 2021).

Second, the scatter around the TRGB/Cepheid fit to the bTF (shown as the black line)
is reduced by 5\% for the color and B+D models.  This is a promising trend for the
bTF studies as a whole, but indicates diminishing returns even with more accurate
colors.  Point by point color analysis of the disks of spiral would be required, with
more detailed modeling of the SFH of each point, in order to increase the accuracy of
the $\Upsilon_*$ values and, therefore, the final stellar masses.  Increasing the
information of the interior stellar population, for example, by spectral indices,
would improve the model fits and applied $\Upsilon_*$ values.

\section{Summary}

We present a continuation of our color versus mass-to-light ($\Upsilon_*$) studies of
galaxies using stellar population models that include two important changes; 1)
extensions to bluer and redder colors to match the observed range in real galaxies
and 2) a new bulge+disk model to produce more accurate $\Upsilon_*$ for early-type
spirals.   While new stellar masses for late-type galaxies calculated from these new
models differ very little from our original prescriptions, correct application of
these models to the bulge and disk colors of early-type spirals can have factor of
two change in their total stellar masses, which improves the linearity of the
baryonic Tully-Fisher relation on the high mass end.

Our technique differs in several key ways from previous studies.  To summarize:

\begin{itemize}

\item We deduce scenarios for the star formation history and chemical enrichment for
galaxies using the main-sequence and mass-metallicity relations.  Constrained by
optical and near-IR colors, the baseline models predict the $\Upsilon_*$ across all
filters of interest.  The scatter in the main-sequence and mass-metallicity relations
matches the dispersion in colors and provides a measure of uncertainty to the
$\Upsilon_*$ versus color relations.

\item We develop a series of different methods to calculate $\Upsilon_*$ using colors
from different components (i.e., bulge and disk) versus total colors versus simple
galaxy morphology, and confirm their internal consistency across multiple bandpasses.
We confirm the accuracy of our SFH scenarios using the model predictions in two color
space as well as galaxy morphology versus color diagrams.  We find that UV and blue
colors can reduce the uncertainty by 15\% for low mass galaxies with irregular SFHs.
A $\Upsilon_*$ webtool is available for the community at
http://abyss.uoregon.edu/$\sim$js/sfh.

\item We compare our new $\Upsilon_*$ values with our original SPARC stellar masses
and find a negligible difference for low mass galaxies and a slight increase in the
early-type spirals due to the proper treatment of a bulge component.  The slight
increase on the high mass end of the baryonic Tully-Fisher relation improves its
linearity and reduces the error in the slope.

\end{itemize}

We found, in deference to other color versus $\Upsilon_*$ studies, that the near-IR
filters produce the most accurate stellar masses and that further improvements to the
accuracy of $\Upsilon_*$ will require more information of the SFH of individual
galaxies (e.g., through spectral indices).  The uses of our new $\Upsilon_*$ models
reduces the scatter in the baryonic Tully-Fisher relation.  Analysis of the scatter
in the baryonic Tully-Fisher relation suggests that there is no deviation from
linearity in log space.  Reducing scatter further in the bTF will require significant
improvements to distance estimates rather than better galaxy photometry or kinematic
work as the estimates of stellar masses are no longer the limiting parameter in the
bTF.

%\begin{acknowledgements}
\bigskip

Software for this project was developed under NASA's AIRS and ADP
Programs. This work is based in part on observations made with the Spitzer Space
Telescope, which is operated by the Jet Propulsion Laboratory, California Institute
of Technology under a contract with NASA.  Support for this work was provided by NASA
through an award issued by JPL/Caltech. Other aspects of this work were supported in
part by NASA ADAP grant NNX11AF89G and NSF grant AST 0908370. As usual, this research has
made use of the NASA/IPAC Extragalactic Database (NED) which is operated by the Jet
Propulsion Laboratory, California Institute of Technology, under contract with
the National Aeronautics and Space Administration.

%\end{acknowledgements}


\begin{thebibliography}{99}

\bibitem[Angus et al.(2016)]{2016A&A...585A..17A} Angus, G.~W., Gentile, G., \& Famaey, B.\ 2016, \aap, 585, A17. doi:10.1051/0004-6361/201527122
\bibitem[Aniyan et al.(2021)]{2021MNRAS.500.3579A} Aniyan, S., Ponomareva, A.~A., Freeman, K.~C., et al.\ 2021, \mnras, 500, 3579.
\bibitem[Bershady et al.(2010)]{2010ApJ...716..198B} Bershady, M.~A., Verheijen, M.~A.~W., Swaters, R.~A., et al.\ 2010, \apj, 716, 198.
\bibitem[Blanton et al.(2017)]{2017AJ....154...28B} Blanton, M.~R., Bershady, M.~A., Abolfathi, B., et al.\ 2017, \aj, 154, 28. doi:10.3847/1538-3881/aa7567
\bibitem[Brouwer et al.(2021)]{2021A&A...650A.113B} Brouwer, M.~M., Oman, K.~A., Valentijn, E.~A., et al.\ 2021, \aap, 650, A113. doi:10.1051/0004-6361/202040108
\bibitem[Buta et al.(1994)]{1994AJ....107..118B} Buta, R., Mitra, S., de Vaucouleurs, G., et al.\ 1994, \aj, 107, 118. doi:10.1086/116838
\bibitem[Calura \& Menci(2011)]{2011MNRAS.413L...1C} Calura, F. \& Menci, N.\ 2011, \mnras, 413, L1. doi:10.1111/j.1745-3933.2011.01017.x
\bibitem[Conroy \& Gunn(2010)]{2010ApJ...712..833C} Conroy, C. \& Gunn, J.~E.\ 2010, \apj, 712, 833. doi:10.1088/0004-637X/712/2/833
\bibitem[Cook et al.(2014)]{2014MNRAS.445..899C} Cook, D.~O., Dale, D.~A., Johnson,
B.~D., et al.\ 2014, \mnras, 445, 899
\bibitem[Costantin et al.(2021)]{2021ApJ...913..125C} Costantin, L., P{\'e}rez-Gonz{\'a}lez, P.~G., M{\'e}ndez-Abreu, J., et al.\ 2021, \apj, 913, 125. doi:10.3847/1538-4357/abef72
\bibitem[Cresci et al.(2019)]{2019A&A...627A..42C} Cresci, G., Mannucci, F., \& Curti, M.\ 2019, \aap, 627, A42. doi:10.1051/0004-6361/201834637
\bibitem[Di Teodoro et al.(2021)]{2021MNRAS.507.5820D} Di Teodoro, E.~M., Posti, L.,
Ogle, P.~M., et al.\ 2021, \mnras, 507, 5820. doi:10.1093/mnras/stab2549
\bibitem[Eftekhari et al.(2021)]{2021MNRAS.504.2190E} Eftekhari, E., Vazdekis, A., \& La Barbera, F.\ 2021, \mnras, 504, 2190. doi:10.1093/mnras/stab976
\bibitem[Ge et al.(2019)]{2019MNRAS.485.1675G} Ge, J., Mao, S., Lu, Y., et al.\ 2019, \mnras, 485, 1675. doi:10.1093/mnras/stz418
\bibitem[Ge et al.(2021)]{2021MNRAS.507.2488G} Ge, J., Mao, S., Lu, Y., et al.\ 2021, \mnras, 507, 2488.  doi:10.1093/mnras/stab2341
\bibitem[Gil de Paz \& Madore(2005)]{2005ApJS..156..345G} Gil de Paz, A. \& Madore, B.~F.\ 2005, \apjs, 156, 345.
doi:10.1086/427068
\bibitem[Graham \& Worley(2008)]{2008MNRAS.388.1708G} Graham, A.~W. \& Worley, C.~C.\ 2008, \mnras, 388, 1708. doi:10.1111/j.1365-2966.2008.13506.x
\bibitem[Gavazzi et al.(2002)]{2002ApJ...576..135G} Gavazzi, G., Bonfanti, C., Sanvito, G., et al.\ 2002, \apj, 576, 135. 
\bibitem[Into \& Portinari(2013)]{2013MNRAS.430.2715I} Into, T. \& Portinari, L.\ 2013, \mnras, 430, 2715. doi:10.1093/mnras/stt071
\bibitem[Johnston et al.(2014)]{2014MNRAS.441..333J} Johnston, E.~J., Arag{\'o}n-Salamanca, A., \& Merrifield, M.~R.\ 2014, \mnras, 441, 333.  doi:10.1093/mnras/stu582
\bibitem[Kennedy et al.(2016)]{2016A&A...593A..84K} Kennedy, R., Bamford, S.~P., H{\"a}u{\ss}ler, B., et al.\ 2016, \aap, 593, A84. doi:10.1051/0004-6361/201628715
\bibitem[Kewley \& Ellison(2008)]{2008ApJ...681.1183K} Kewley, L.~J. \& Ellison, S.~L.\ 2008, \apj, 681, 1183. doi:10.1086/587500
\bibitem[Lelli et al.(2016)]{2016AJ....152..157L} Lelli, F., McGaugh, S.~S., \& Schombert, J.~M.\ 2016, \aj, 152, 157. doi:10.3847/0004-6256/152/6/157
\bibitem[Lelli et al.(2019)]{2019MNRAS.484.3267L} Lelli, F., McGaugh, S.~S., Schombert, J.~M., et al.\ 2019, \mnras, 484, 3267. 
\bibitem[Li et al.(2018)]{2018MNRAS.476.1765L} Li, H., Mao, S., Cappellari, M., et al.\ 2018, \mnras, 476, 1765.  doi:10.1093/mnras/sty334
\bibitem[Lower et al.(2020)]{2020MNRAS.494..228L} Lower, M.~E., Bailes, M., Shannon, R.~M., et al.\ 2020, \mnras, 494, 228. 
\bibitem[MacArthur et al.(2008)]{2008ApJ...680...70M} MacArthur, L.~A., Ellis, R.~S., Treu, T., et al.\ 2008, \apj, 680, 70. doi:10.1086/587887
\bibitem[Maiolino \& Mannucci(2019)]{2019A&ARv..27....3M} Maiolino, R. \& Mannucci, F.\ 2019, \aapr, 27, 3. doi:10.1007/s00159-018-0112-2
\bibitem[Martinsson et al.(2013)]{2013A&A...557A.131M} Martinsson, T.~P.~K., Verheijen, M.~A.~W., Westfall, K.~B., et al.\ 2013, \aap, 557, A131. 
\bibitem[McGaugh et al.(2017)]{2017ApJ...851...22M} McGaugh, S.~S., Schombert, J.~M., \& Lelli, F.\ 2017, \apj, 851, 22. doi:10.3847/1538-4357/aa9790
\bibitem[McGaugh et al.(2020)]{2020RNAAS...4...45M} McGaugh, S.~S., Lelli, F., \& Schombert, J.~M.\ 2020, Research Notes of the American Astronomical Society, 4, 45. 
\bibitem[McQuinn et al.(2010)]{2010ApJ...724...49M} McQuinn, K.~B.~W., Skillman, E.~D., Cannon, J.~M., et al.\ 2010, \apj, 724, 49. doi:10.1088/0004-637X/724/1/49
\bibitem[Moster et al.(2013)]{2013MNRAS.428.3121M} Moster, B.~P., Naab, T., \& White, S.~D.~M.\ 2013, \mnras, 428, 3121. doi:10.1093/mnras/sts261
\bibitem[Peterken et al.(2021)]{2021MNRAS.502.3128P} Peterken, T., Arag{\'o}n-Salamanca, A., Merrifield, M., et al.\ 2021, \mnras, 502, 3128. 
\bibitem[Portinari et al.(2004)]{2004MNRAS.347..691P} Portinari, L., Sommer-Larsen, J., \& Tantalo, R.\ 2004, \mnras, 347, 691. doi:10.1111/j.1365-2966.2004.07207.x
\bibitem[Prantzos \& Boissier(2000)]{2000MNRAS.313..338P} Prantzos, N. \& Boissier, S.\ 2000, \mnras, 313, 338. 
\bibitem[Roediger \& Courteau(2015)]{2015MNRAS.452.3209R} Roediger, J.~C. \& Courteau, S.\ 2015, \mnras, 452, 3209. doi:10.1093/mnras/stv1499
\bibitem[Samland \& Gerhard(2003)]{2003A&A...399..961S} Samland, M. \& Gerhard, O.~E.\ 2003, \aap, 399, 961. doi:10.1051/0004-6361:20021842
\bibitem[Sandage \& Visvanathan(1978)]{1978ApJ...223..707S} Sandage, A. \& Visvanathan, N.\ 1978, \apj, 223, 707. doi:10.1086/156305
\bibitem[Sandage \& Tammann(1983)]{1983C&T....99R..63S} Sandage, A. \& Tammann, G.~A.\ 1983, Ciel et Terre, 99, 63
\bibitem[Schombert \& Rakos(2009)]{2009AJ....137..528S} Schombert, J. \& Rakos, K.\ 2009, \aj, 137, 528. doi:10.1088/0004-6256/137/1/528
\bibitem[Schombert et al.(2013)]{2013AJ....146...41S} Schombert, J., McGaugh, S., \& Maciel, T.\ 2013, \aj, 146, 41. doi:10.1088/0004-6256/146/2/41
\bibitem[Schombert \& McGaugh(2014)]{2014PASA...31...36S} Schombert, J. \& McGaugh, S.\ 2014, \pasa, 31, e036. doi:10.1017/pasa.2014.32
\bibitem[Schombert \& McGaugh(2015)]{2015AJ....150...72S} Schombert, J. \& McGaugh, S.\ 2015, \aj, 150, 72. doi:10.1088/0004-6256/150/3/72
\bibitem[Schombert(2016)]{2016AJ....152..214S} Schombert, J.~M.\ 2016, \aj, 152, 214. doi:10.3847/0004-6256/152/6/214
\bibitem[Schombert et al.(2019)]{2019MNRAS.483.1496S} Schombert, J., McGaugh, S., \& Lelli, F.\ 2019, \mnras, 483, 1496. doi:10.1093/mnras/sty3223
\bibitem[Schombert et al.(2020)]{2020AJ....160...71S} Schombert, J., McGaugh, S., \& Lelli, F.\ 2020, \aj, 160, 71. doi:10.3847/1538-3881/ab9d88
\bibitem[Schombert et al.(2020)]{2020AJ....160...71S} Schombert, J., McGaugh, S., \& Lelli, F.\ 2020, \aj, 160, 71. doi:10.3847/1538-3881/ab9d88
\bibitem[Schombert \& McGaugh(2021)]{2021AJ....161...91S} Schombert, J. \& McGaugh, S.\ 2021, \aj, 161, 91. doi:10.3847/1538-3881/abd54d
\bibitem[Sheth et al.(2010)]{2010PASP..122.1397S} Sheth, K., Regan, M., Hinz, J.~L., et al.\ 2010, \pasp, 122, 1397
\bibitem[Speagle et al.(2014)]{2014ApJS..214...15S} Speagle, J.~S., Steinhardt, C.~L., Capak, P.~L., et al.\ 2014, \apjs, 214, 15. doi:10.1088/0067-0049/214/2/15
\bibitem[Stone et al.(2021)]{2021MNRAS.508.1870S} Stone, C.~J., Arora, N., Courteau, S., et al.\ 2021, \mnras, 508, 1870.  doi:10.1093/mnras/stab2709
\bibitem[Tasca \& White(2011)]{2011A&A...530A.106T} Tasca, L.~A.~M. \& White, S.~D.~M.\ 2011, \aap, 530, A106. 
\bibitem[Taylor et al.(2011)]{2011MNRAS.418.1587T} Taylor, E.~N., Hopkins, A.~M., Baldry, I.~K., et al.\ 2011, \mnras, 418, 1587. doi:10.1111/j.1365-2966.2011.19536.x
\bibitem[Thomas et al.(2005)]{2005ApJ...621..673T} Thomas, D., Maraston, C., Bender, R., et al.\ 2005, \apj, 621, 673. doi:10.1086/426932
\bibitem[Tonini et al.(2016)]{2016MNRAS.459.4109T} Tonini, C., Mutch, S.~J., Croton, D.~J., et al.\ 2016, \mnras, 459, 4109. doi:10.1093/mnras/stw956
\bibitem[Tully et al.(2016)]{2016AJ....152...50T} Tully, R.~B., Courtois, H.~M., \& Sorce, J.~G.\ 2016, \aj, 152, 50. doi:10.3847/0004-6256/152/2/50
\bibitem[Weisz et al.(2011)]{2011ApJ...739....5W} Weisz, D.~R., Dalcanton, J.~J., Williams, B.~F., et al.\ 2011, \apj, 739, 5. doi:10.1088/0004-637X/739/1/5
\bibitem[Weldon et al.(2020)]{2020MNRAS.491.2254W} Weldon, A., Ly, C., \& Cooper, M.\ 2020, \mnras, 491, 2254. doi:10.1093/mnras/stz3047
\bibitem[Zibetti et al.(2009)]{2009MNRAS.400.1181Z} Zibetti, S., Charlot, S., \& Rix, H.-W.\ 2009, \mnras, 400, 1181. doi:10.1111/j.1365-2966.2009.15528.x
\bibitem[van den Bosch(1998)]{1998ApJ...507..601V} van den Bosch, F.~C.\ 1998, \apj, 507, 601. doi:10.1086/306354
\bibitem[Verheijen(2001)]{2001ApJ...563..694V} Verheijen, M.~A.~W.\ 2001, \apj, 563, 694. 
\bibitem[Y{\i}ld{\i}z et al.(2017)]{2017MNRAS.464..329Y} Y{\i}ld{\i}z, M.~K., Serra,
P., Peletier, R.~F., et al.\ 2017, \mnras, 464, 329. doi:10.1093/mnras/stw2294


\end{thebibliography}
\end{document}